\documentstyle[psfig,twocolumn,prl,aps,amssymb]{revtex}
\begin{document}
\wideabs{
\draft

\title{Kondo effect of non-magnetic impurities 
and the co-existing charge order
in the cuprate superconductors}
\author{Kwon Park}
\address{Department of Physics, Yale University, P.O. Box 208120,
New Haven, CT 06520-8120}
\date{\today}

\maketitle

\begin{abstract}
We present a theory of Kondo effect 
caused by an induced magnetic moment near 
non-magnetic impurities such as Zn and Li
in the cuprate superconductors. 
Based on the co-existence of charge order and superconductivity,
a natural description of the induced moment 
and the resulting Kondo effect
is obtained
in the framework of bond-operator theory
of microscopic $t$-$J$-$V$ Hamiltonian.
The local density of state near impurities
is computed in a self-consistent Bogoliubov-de Gennes theory which
shows a low-energy peak in the middle of superconducting gap.
Our theory also suggests that the charge order can be enhanced
near impuries. 
\end{abstract}

}

Impurities in the cuprate superconductors have attracted 
much interest because of their role 
as an effective tool in probing 
the local electronic and magnetic properties of 
high $T_c$ superconductors.
Especially remarkable have been
recent experiments in scanning tunneling microscopy (STM) 
\cite{Davis,Yazdani}
of surfaces of the cuprate superconductor 
Bi$_2$Sr$_2$CaCu$_2$O$_{8+\delta}$ (BSCCO)
which showed a low-bias peak in the differential tunneling conductance
in the vicinity of Zn impurities \cite{Davis}. 
Since the Zn$^{++}$ ion has spin $S=0$, it seems natural to view
each Zn site as just a potential scatterer, 
and therefore to interpret the peak as 
a quasibound state \cite{Balatsky}.
However, there are some experimental features which cannot
be naturally explained by this theory \cite{Polkovnikov}.

\begin{figure}
\centerline{\psfig{figure=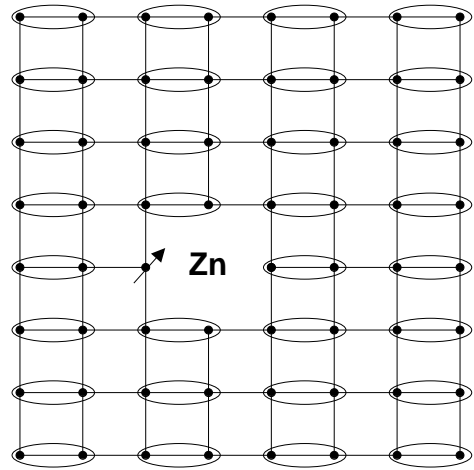,width=3.2in,angle=0}}
\caption{Schematic diagram of the ground state of doped antiferromagnet 
with Zn impurity, which is assumed to have 
bond-centered charge order with period of two lattice spacing
(columnal spin-Peierls order). 
Ellipses denote the valence bonds formed by two neighboring spins:
$\frac{1}{\sqrt{2}}(|\uparrow,\downarrow\rangle - |\downarrow,\uparrow\rangle)$.
The arrow near Zn impurity indicates the induced spin moment.
Note that a similar picture of moment formation applies to other charge order states.
\label{fig1}}
\end{figure}

More importantly, a series of NMR experiments \cite{Bobroff,NMR} 
have shown clear evidences that there is an induced magnetic moment 
with spin $S=1/2$ near Zn impurities.
Therefore, it is imperative for a consistent theory to address 
the problem of impurity in the framework of Kondo effect.
Though there have been several, previous works based on 
the Kondo physics \cite{Polkovnikov,ZhuTing},  
their approach was based on the {\it ad hoc} assumptions 
on the existence of induced spin moment near Zn impurities. 
Especially, the location of induced spin moment
and the range of Kondo interaction
were chosen in {\it ad hoc} manner.
In this article, we would like to provide 
a self-consistent theory of Kondo effect 
based on the co-existence of charge order 
and superconductivity. 
We take this assumption to be natural because 
one of the important consequences of 
charge order is the induced $S=1/2$ moment
near non-magnetic impurities \cite{Bond}, 
as schematically shown in Fig.\ref{fig1}.

Much more direct evidences 
for the relevance of charge order
in the high $T_c$ superconductors 
have been recently obtained in 
several remarkable STM experiments, 
one of which was performed by Hoffman {\it et al.} \cite{Hoffman} 
where the charge order  
with a period of four lattice spacing was observed near 
vortex cores in BSCCO.
As predicted in Ref.\cite{Bond}, the superconductivity 
is locally suppressed near the cores of vortices, and 
the static charge order appears in such regions.
Another STM measurement in optimally doped BSCCO 
was performed by Howald {\it et al.} \cite{Howald}, 
showing clear evidences for
the static charge order {\it even in zero magnetic field}. 
In addition to the STM measurements, 
inelastic neutron scattering techniques 
have been also used to provide important
evidences for the charge order
in La$_{2-x}$Sr$_x$CuO$_4$ (LSCO) 
and YBa$_2$Cu$_3$O$_{6+x}$ (YBCO) \cite{MookCDW,McQueeney}.

Note our viewpoint that, though the charge order may not be static 
in all regions of superconducting phase, 
the time scale of charge order fluctuation is long enough
to ensure the superconducting correlation and the Kondo spin physics.
Therefore, in the rest of discussions,
we will assume the static charge order 
with period of two lattice spacing (the spin-Peierls order), 
which, we believe, captures the essential physics
in spite of its simplicity.  
Also, remember that the charge orders with different period
are most likely to coexist because, as shown in Ref.\cite{Vojta},
the periodicity is rather sensitive to various parameters
including the doping concentration. 
%the ratio $t/J$ and 
%the strength of the nearest-neighbor Coulomb interaction, $V$.

Recently, 
a natural theoretical framework 
was developed to describe the co-existing phase of
charge density wave (CDW) order and superconductivity,
based on the bond-operator representation \cite{Bond}. 
It was shown that the saddle-point approximation of 
this bond-operator theory
consistently interpolates
the phase of Mott insulator at low dopings and
that of superconductor with nodal fermions at moderate dopings. 
Therefore, it would be very interesting to explore the question whether
the bond-operator theory gives rise to 
the low-energy peak of local density of state 
in the vicinity of non-magnetic impurities
without {\it ad hoc} assumptions.
It would be also interesting to study the effect of impurity
on the charge order.

Now we begin our bond-operator theory by setting up the 
exact mapping between the bond operators and the usual electron
creation operators \cite{Bond}. 
Let $c^{\dagger}_{1a}$ and $c^{\dagger}_{2a}$ $(a= \uparrow,\downarrow)$
be the electron creation operators on the two sites of a pair.
When we project out all states with two electrons at the same site, 
the electronic Hilbert space for a pair of sites is composed of
nine states which can be expressed in terms of 
the ``bond particle'' creation operators defined by:
\begin{eqnarray}
s^{\dagger} |v\rangle &=& \frac{1}{\sqrt{2}} \varepsilon_{ab}
c^{\dagger}_{1a} c^{\dagger}_{2b} |0\rangle, \\
t^{\dagger}_{\alpha} |v\rangle &=& \frac{1}{\sqrt{2}} 
\sigma^{\alpha}_{bc} \varepsilon_{ca} 
c^{\dagger}_{1a} c^{\dagger}_{2b} |0\rangle, \\
h^{\dagger}_{1a} |v\rangle &=& c^{\dagger}_{1a} |0\rangle, \\
h^{\dagger}_{2a} |v\rangle &=& c^{\dagger}_{2a} |0\rangle, \\
d^{\dagger} |v\rangle &=& |0\rangle 
\end{eqnarray}
where $|0\rangle$ is the electron vacuum and 
$|v\rangle$ is an imaginary vacuum void of any bond particles.
Remember that $\sigma^{\alpha}_{ab}$ $(\alpha=x,y,z)$ 
are the Pauli matrices, and 
$\varepsilon_{ab}$ is the second-rank antisymmetric tensor with
$\varepsilon_{\uparrow\downarrow}=+1$.
The opertators $s,d,t_{\alpha}$ all obey the canonical
boson commutation relations, while the $h_{1a}, h_{2a}$ obey
the canonical fermion relations.
Since the total Hilbert space of these five bosons and four fermions
is much larger than that of the physical nine states,
we must impose the following constraint on the bond particle Hilbert space.
\begin{equation}
s^{\dagger}s +t^{\dagger}_{\alpha}t_{\alpha}
+h^{\dagger}_{1a}h_{1a} +h^{\dagger}_{2a}h_{2a}
+d^{\dagger}d = 1.
\label{constraint}
\end{equation}
In the subspace contrained by Eq. (\ref{constraint}) we can
write the exact expressions for electron operators in terms
of bond operators:
\begin{eqnarray}
c^{\dagger}_{1a} &=& h^{\dagger}_{1a}d +\frac{1}{\sqrt{2}}\varepsilon_{ab}
s^{\dagger}h_{2b}-\frac{1}{\sqrt{2}}\varepsilon_{ac}\sigma^{\alpha}_{cb}
t^{\dagger}_{\alpha}h_{2b},\\
c^{\dagger}_{2a} &=& h^{\dagger}_{2a}d +\frac{1}{\sqrt{2}}\varepsilon_{ab}
s^{\dagger}h_{1b}+\frac{1}{\sqrt{2}}\varepsilon_{ac}\sigma^{\alpha}_{cb}
t^{\dagger}_{\alpha}h_{1b},\\
S_{1\alpha} &=& \frac{1}{2}(s^{\dagger}t_{\alpha}
+t^{\dagger}_{\alpha}s
-\epsilon_{\alpha\beta\gamma}t^{\dagger}_{\beta}t_{\gamma})
+\frac{1}{2}\sigma^{\alpha}_{ab}h^{\dagger}_{1a}h_{1b},\\
S_{2\alpha} &=& -\frac{1}{2}(s^{\dagger}t_{\alpha}
+t^{\dagger}_{\alpha}s
+\epsilon_{\alpha\beta\gamma}t^{\dagger}_{\beta}t_{\gamma})
+\frac{1}{2}\sigma^{\alpha}_{ab}h^{\dagger}_{2a}h_{2b},
\end{eqnarray}
where $\epsilon_{\alpha\beta\gamma}$ is the third-rank antisymmetric
tensor with $\epsilon_{xyz}=+1$, and, as usual, the electon spin operator 
is defined as $S_{\alpha}= 1/2 c^{\dagger}_a\sigma^{\alpha}_{ab}c_b$.

Equiped with the bond-operator representation of electron operators,
we now apply the Bogoliubov-de Gennes (BdG) theory 
to solve the $t$-$J$-$V$ model in finite systems
containing a single impurity with (infinitely) strong repulsion, $U$,
which is defined by:
\begin{eqnarray}
H &=& -t\sum_{\langle i,j \rangle} (c^{\dagger}_{ia}c_{ja}
+c^{\dagger}_{ja}c_{ia}) 
+J\sum_{\langle i,j \rangle} S_{i\alpha}S_{j\alpha} 
\nonumber \\
&+&V\sum_{\langle i,j \rangle}c^{\dagger}_{ia}c_{ia}c^{\dagger}_{jb}c_{jb}
-\mu \sum_i c^{\dagger}_{ia}c_{ia}
+U c^{\dagger}_{i_0 a}c_{i_0 a}
\end{eqnarray}
where $i_0$ denotes the position of impurity, 
and $\mu$ is the chemical potential. 
Also, it is implicitly
assumed that all states with two electrons on any site have been 
projected out. 
Written in terms of bond operators, 
the Hamiltonian is physically meaningful  
only when the contraint condition in Eq.(\ref{constraint})
is satisfied simultaneously.
In this article,
as usual in the mean field theory,
only the average of this constraint will be satisfied
via the Lagrange multiplier method. Note, however, that
the average is taken over quantum fluctuations, not over space, 
so that the constraint will be satisfied individually at each site.

While general techniques of the bond-operator method
can be found in Ref.\cite{Details} in detail,
several conceptial and technical points are worth mentioning, 
especially related to the impurity problem.
(1) The nearst-neighbor Coulomb repulsion, $V$, is not only
physically reasonable, but also gives rise to 
an important consequence. Without $V$, the pairing of holes
primarily occurs through the condensation of $d$-bosons, which
results in a very short-range $s$-wave-like pairing.
It is only when the $d$-boson condensation is suppressed
by a large $V$ that the superconducting state develops
$d$-wave-like pairing, and nodal fermions emerge. 
Remember that we do not make any assumptions 
either on the emergence of superconducting state or
the symmetry of pairing. They are obtained as a natural consequence
of our saddle-point bond-operator theory of $t$-$J$-$V$ model
at moderate dopings.

(2) The chemical potential, $\mu$, should be determined by fixing
the average hole concentration, $x$, in the region far away from the impurity. 
In our finite system, we first obtain the value of $\mu$ without
introducing the impurity, and then use it for the case of impurity
by assuming that the system size is large enough so that a single
impurity does not change the chemical potential.

(3) The impurity potential, $U$, is taken to be infinitely repulsive so that
electrons are completely depleted from the Zn site. 
In the bond-operator formalism, one of the $h$-fermions, 
say $h_{1a} (a=\uparrow,\downarrow)$, is
pinned at the unpaired site near the Zn impurity.
(See Fig.\ref{fig1}.) 
Of course, it is possible without charge order
in the doped antiferromagnets
that the electron escapes from the unpaired site all together, 
leaving the vacancy instead of lone spin moment.
It is assumed, however, that the empty state is energetically unfavorable
because of the similar reason why $d$-boson condensation is suppressed 
for large $V$. 

(4) Finally, unlike the previous work in Ref.\cite{ZhuTing},
there are now two sets of BdG equations both for 
the bosons ($t_{\alpha}$) and the fermions ($h_{1a}$ and $h_{2a}$).
Consequently, the number of self-consistency conditions
for the normal and anomalous exchange energies
is greatly increased, 
which limits the system size significantly.
The number of self-consistence equations
is $11\times N^2 /2$ with $N^2$ being the number of sites. 
Also, note that the diagonalization of boson BdG matrix 
is not as straightforward as that of fermion BdG matrix.
It is performed by using the method in Ref.\cite{BosonDiag}.

\begin{figure}
\centerline{\psfig{figure=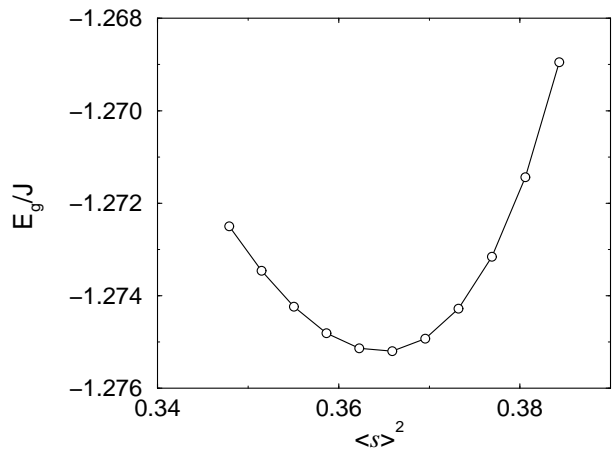,width=3.8in,angle=-90}}
\caption{Ground state energy as a function of the condensation
density of $s$-boson for the sqaure lattice system of $8\times8$ sites
without the impurity.
Here $t/J = 1.5$ and the hole concentration $x=0.3$.
%Note that the $s$-boson condensation density can be viewed as 
%an order parameter for the spin-Peierls order.
\label{fig2}}
\end{figure}

The saddle-point theory is not yet complete  
alone with the aforementioned self-consistent BdG equations, 
but it still needs to determine 
the condensation density of $s$-boson, $\langle s {\rangle}^2$,
by minimizing the ground state energy.
%with the constraint in Eq.(\ref{constraint}) satisfied simultaneously.
Similar to the chemical potential, the Lagrange multiplier
associated with the constraint is first computed as a function of 
$\langle s \rangle^2$ 
for the situation when there is no impurity,
by satisfying the constraint in the mean field level. 
Then, $\langle s \rangle^2$ is fixed to be the value minimizing 
the ground state energy, 
at which point the Lagrange multiplier is also fixed.
Fig.\ref{fig2} shows the ground state energy as a function of
$s$-boson condensation density which also can serve as 
an order parameter for the spin-Peierls order.
%As one can see in Fig.\ref{fig2}, the ground state energy
%is minimized around at $\langle s \rangle^2 = 0.365$ which
%shows a spontaneous translational symmetry breaking due to
%the formation of spin-Peierls order.
After the impurity is introduced, 
the position-dependent condensation density, $\langle s_i {\rangle}^2$,
is determined in turn
by satisfying the constraint at each site
with the Lagrange multiplier determined previously.
In this way, it is guaranteed (at least in the mean field level)
that the ground state energy is minimized
and also the constraint is satisfied at the same time.

Now let us turn to the physical observables.
As mentioned in the introduction,
the local density of state (LDOS) 
can be obtained directly in the STM experiments
through the measurement of differential tunneling conductance,
which show a low-energy peak in LDOS near non-magnectic impurities
while the superconducting coherence peak is 
greatly suppressed \cite{Davis}.
Theoretically, the local density of state can be computed 
in terms of the spectral function given by:
\begin{equation}
\rho({\bf r},\omega) = \;$Im$\; G^{ret}({\bf r},\omega +i\delta),
\end{equation}
where the retarded electronic Green function,   
$G^{ret}({\bf r},\omega +i\delta)$, is obtained by
using the usual analytic continuation from
the time-ordered electronic Green function 
$G({\bf r},\tau)=
\langle T_{\tau} c_{a}({\bf r},\tau) c_{a}^{\dagger} ({\bf r},0) \rangle $
$(a=\uparrow,\downarrow)$.
In the bond-operator formalism, 
the spectral function of electron on the site 1 
of dimer located at ${\bf r}_i$
is related to that of $h_2$-fermion via:
\begin{equation}
\rho_{1}({\bf r}_i,\omega) \cong -\frac{1}{2} \langle s_i {\rangle}^2
\;$Im$\; G^{ret}_{h_2}({\bf r}_i,-\omega -i\delta),
\label{dos}
\end{equation}
where 
$G_{h_2}({\bf r}_i,\tau)=
\langle T_{\tau} h_{2a}({\bf r}_i,\tau) 
h_{2a}^{\dagger} ({\bf r}_i,0) \rangle$. 
The electronic spectral function at site 2 is related to that
of $h_1$-fermion in similar way.
In Eq.(\ref{dos}) the approximation is made when 
the contribution from magnons ($t_{\alpha}$) is ignored
because they are high-energy modes.

\begin{figure}
\centerline{\psfig{figure=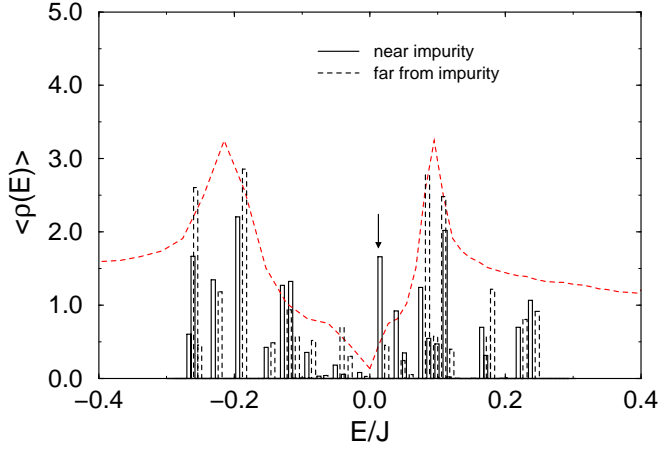,width=4.0in,angle=-90}}
\caption{Local density of state (LDOS) near the impurity (solid histograms)
and far away from the impurity (dashed histograms) as a function of energy
in the finite square-lattice system of $12\times12$ sites
with $t/J = 1.5$ and $x=0.3$.
%(For the precise definition of 
%``near impurity'' and ``far away from impurity'',
%refer to the text.)
The dashed line is just a guide to eye for the
LDOS far away from impurity, whose shape is obtained 
from the analytic computation for the uniform system, 
and is fitted roughly to follow the LDOS computed in finite system.
Remember that the LDOS is not normalized due to the discrete nature
of energy spectrum in finite system.
It is important to note 
that the low-energy peak (denoted by arrow)
develops in LDOS near the impurity, while
the coherence peak is suppressed, 
which is in a good agreement with experiments.
\label{fig3}}
\end{figure}

To compare the theoretical LDOS with 
the differential tunneling conductance measured in STM experiments,
we need to take into account some key properties of 
realistic surface structure of BSCCO.  
Between the superconducting CuO$_2$ layer and the STM tip, there is always
the BiO layer formed in such a way that each Bi atom is located directly 
above each Cu or Zn atom. 
So it is reasonable to assume that
the Bi atom will block tunneling currents from 
reaching the Cu/Zn atom directly below the STM tip, 
and so the STM measures the LDOS contributed
by the four nearest-neighboring sites 
instead of the single site directly below the tip.
\cite{ZhuTing,Zaanen}.
Under this assumption, 
the LDOS measured in STM experiments, $\langle\rho({\bf r}_i,\omega)\rangle$, 
may be identified
with an average of LDOS from the nearest-neighboring
sites: $\langle\rho({\bf r}_i,\omega)\rangle \propto 
\rho({\bf r}_i +\hat{x},\omega)
+\rho({\bf r}_i -\hat{x},\omega)
+\rho({\bf r}_i +\hat{y},\omega)
+\rho({\bf r}_i -\hat{y},\omega)$.
Fig.\ref{fig3} shows $\langle\rho({\bf r}_i,E)\rangle$ 
near and far away from the Zn impurity
as a function of energy $E/J$, where
the LDOS near impurity is defined as 
$\langle\rho({\bf r}_i,\omega)\rangle$ with
${\bf r}_i$ indicating the position of the Zn impurity.
Compared to the LDOS far away from the impurity,
the low-energy peak (denoted by the arrow in graph)
develops near the Zn impurity, while
the coherence peak is somewhat reduced 
indicating the suppression of superconductivity, 
which is in a reasonably good agreement with STM measurements.

\begin{figure}
\centerline{\psfig{figure=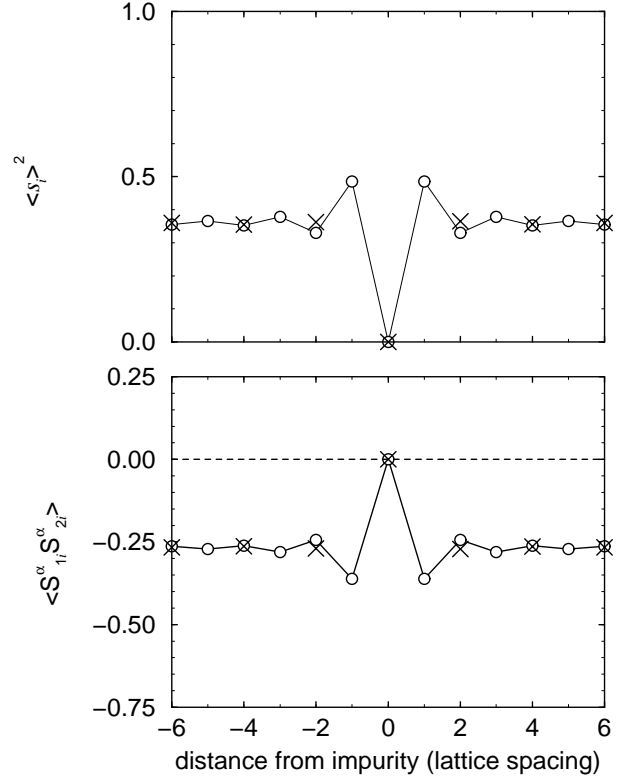,width=3.7in,angle=0}}
\caption{Condensation density of $s$-boson, 
$\langle s_i \rangle^2$ (top panel), and
the equal-time spin-spin correlation within dimer,
$\langle S^{\alpha}_{1i} S^{\alpha}_{2i} \rangle$, (bottom panel)
as a function of distance from the impurity
in the finite square-lattice system of $12\times12$ sites
with $t/J = 1.5$ and $x=0.3$.
Remember that in the bond-operator formalism
$S^{\alpha}_{1i} S^{\alpha}_{2i}= -\frac{3}{4} s_i^2 
+\frac{1}{4} t^{\dagger}_{i} t_{i}$ whose expectation value 
equals to $-3/4$ for the pure singlet valence bond, 
and $1/4$ for the pure triplet magnon. 
Of course, both the condensation density and the correlation are
exactly zero at the impurity site since the Zn impurity is assumed to
avoid electrons completely.
Circles (crosses) in the graph 
are associated with the direction parallel (perpendicular) to
the columnal spin-Peierls order.
It is interesting to observe that 
the spin-Peierls order, or the charge density wave, 
is somewhat enhanced near the impurity
while overall it is not affected much by the presence of impurity
after one lattice spacing. 
\label{fig4}}
\end{figure}

Another physical observable in which we are interested is
the order parameter of charge order,
or the spin-Peierls order parameter, $\langle s_i {\rangle}^2$. 
One of the main questions that we would like to answer in this article
is how much the charge order is affected by the impurity.
It is completely conceivable 
that the spin-Peierls order is locally reduced near impurity 
at the same time when the superconductivity is also suppressed.
In the extreme case the spin-Peierls order can be completely
destroyed while the $S=1$ excitons start to condense locally so that
the magnetic order emerges near the impurity: 
in the bond-operator formalism, 
$\langle t_{\alpha}({\bf r}_i) \rangle \ne 0$
with ${\bf r}_i$ indicating the neighboring sites of impurity.
If so, it may imply the emergence of the spin density wave (SDW)
near impurities.
However, the $\langle s_i {\rangle}^2$
plotted as a function of the distance from impurity
in Fig.\ref{fig4} 
shows the opposite behavior that
overall the spin-Peierls order is not affected much 
by the presence of impurity after one lattice spacing.
In fact, the spin-Peierls order is somewhat 
enhanced at the nearest-neighboring Cu site of the Zn impurity, 
which may suggest an interesting prediction that
the static charge can be observed near the Zn impurities
by using the STM techniques. 
Finally, note that the bottom panel of Fig.\ref{fig4} shows
the equal-time spin-spin correlation within dimer,
$S^{\alpha}_{1i} S^{\alpha}_{2i}= -\frac{3}{4} s_i^2 
+\frac{1}{4} t^{\dagger}_{i} t_{i}$, which suggests
basically the similar physics by showing 
that the effect of the spin $S=1$ excitons is small.

In conclusion, we have applied the bond-operator method 
to address the problem of non-magnetic impurity such as
Zn/Li in the high-$T_c$ cuprate superconductors.
It is shown that, without {\it ad hoc} assumptions,
the low-energy peak of local density of state in the vicinity of impurity
can be computed in the framework of bond-operator formalism. Also,
it is predicted that the charge-density-wave order can be enhanced
near Zn impurities.

The author is indebted to Subir Sachdev 
for very useful discussions and encouragement. 
This research was supported in part 
by US NSF Grant DMR 0098226.

\begin{references}


\bibitem{Davis} E.W. Hudson, S.H. Pan, A.K. Gupta, K.-W. Ng, and J.C. Davis,
Science {\bf 285}, 88 (1999); S.H. Pan, E.W. Hudson, K.M. Lang, H Eisaki, 
S. Uchida, and J.C. Davis, Nature (London) {\bf 403}, 746 (2000).


\bibitem{Yazdani} A. Yazdani, C.M. Howald, C.P. Lutz, a. Kapitulnik, and
D.M. Eigler, Phys. Rev. Lett. {\bf 83}, 176 (1999).



\bibitem{Balatsky} A.V. Balatsky, M.I. Salkola, and A. Rosengren,
Phys. Rev. B {\bf 51}, 15547 (1995); M.I. Salkola, A.V. Balatsky, and
D.J. Scalapino, Phys. Rev. Lett. {\bf 77}, 1841 (1996).


\bibitem{Polkovnikov} A. Polkovnikov, S. Subir, and M. Vojta,
Phys. Rev. Lett. {\bf 86}, 296 (2001).


\bibitem{ZhuTing} Jian-Xin Ting and C.S. Ting, Phys. Rev. B {\bf 64},
060501(R) (2001).




\bibitem{Bobroff} J. Bobroff {\it et al.}, Phys. Rev. Lett. {\bf 83}, 
4381 (1999): J. Bobroff {\it et al.}, Phys. Rev. Lett. {\bf 86}, 4116 (2001).


\bibitem{NMR} H. Alloul {\it et al.}, Phys. Rev. Lett. {\bf 67},
3140 (1991); A.V. Mahajan {\it et al.}, {\it ibid} {\bf 72}, 3100 (1994);
P. Mendels {\it et al.}, Europhys. Lett. {\bf46}, 678 (1999);
M.-H. Julien {\it et al.}, Phys. Rev. Lett. {\bf 84}, 3422 (2000).




\bibitem{Bond} Kwon Park and Subir Sachdev, 
Phys. Rev. B {\bf 64}, 184510 (2001). 



%\bibitem{Rossat} J. Rossat-Mignod {\it et al.}, Physica C 
%{\bf 185-189}, 86 (1991).

%\bibitem{Mook} H.A. Mook {\it et al.}, Phys. Rev. Lett. {\bf 70},
%3490 (1993).

%\bibitem{Fong} H.F. Fong {\it et al.}, Phys. Rev. B {\bf 54},
%6708 (1996): H.F. Fong {\it et al.}, Phys. Rev. Lett. {\bf 78},
%713 (1997).

%\bibitem{He} H. He {\it et al.}, Phys. Rev. Lett. {\bf 86}, 1610 (2001).





\bibitem{Hoffman} J.E. Hoffman, E.W. Hudson, K.M. Lang, V. Madhavan,
H. Eisaki, S. Uchida, and J.C. Davis, Science {\bf 295}, 466 (2002).


\bibitem{Howald} C. Howald, H. Eisaki, N. Kaneko, and A. Kapitulnik,
cond-mat/0201546 (2002).


\bibitem{MookCDW} H.A. Mook, Pengcheng Dai, and F. Dogan,
Phys. Rev. Lett. {\bf 88}, 097004 (2002).



\bibitem{McQueeney} R.J. McQueeney, Y. Petrov, T. Egami, M. Yethiraj,
G. Shirane, and Y. Endoh, Phys. Rev. Lett. {\bf 82}, 628 (1999);
R.J. McQueeney {\it et al.}, Phys. Rev. Lett. {\bf 87}, 077001 (2001); 
R.J. McQueeney {\it et al.}, cond-mat/0105593.



\bibitem{Vojta} Matthias Vojta and Subir Sachdev, Phys. Rev. Lett.
{\bf 83}, 3916 (1999); M. Vojta, Y. Zhang, and S. Sachdev,
Phys. Rev. B {\bf 62}, 6721 (2000).




\bibitem{Details} K. Park and S. Sachdev, cond-mat/0104519 (2001).


\bibitem{BosonDiag} Subir Sachdev, Phys. Rev. B {\bf 45}, 12377 (1992).


\bibitem{Zaanen} I. Martin, A.V. Balatsky, J. Zaanen,
Phys. Rev. Lett. {\bf 88}, 097003 (2002).

\end{references}
\end{document}